\documentclass[3p,times]{elsarticle}

\usepackage{ecrc}

\usepackage{epstopdf}

\volume{00}

\firstpage{1}

\journalname{Nuclear Physics A}

\runauth{M.~Bluhm et al.}

\jid{nupha}

\jnltitlelogo{Nuclear Physics A}

\usepackage{graphicx}
\usepackage{amsmath,amssymb}

\usepackage{enumerate}

\begin{document}

\begin{frontmatter}



\title{Determination of freeze-out conditions from fluctuation observables measured at RHIC}



\author[1]{M.~Bluhm}
\author[2]{P.~Alba}
\author[2]{W.~Alberico}
\author[3]{R.~Bellwied}
\author[2]{V.~Mantovani~Sarti}
\author[4]{M.~Nahrgang}
\author[2]{C.~Ratti}

\address[1]{Department of Physics, North Carolina State University, Raleigh, NC 27695, USA}
\address[2]{Department of Physics, Torino University and INFN, Sezione di Torino, via P. Giuria 1, 10125 Torino, Italy}
\address[3]{Department of Physics, University of Houston, Houston, TX 77204, USA}
\address[4]{Department of Physics, Duke University, Durham, NC 27708, USA}

\begin{abstract}
 We extract chemical freeze-out conditions via a thermal model approach from fluctuation observables measured at RHIC and compare with results from lattice QCD and statistical hadronization model fits. The possible influence of additional critical and non-critical fluctuation sources not accounted for in our analysis is discussed. 
\end{abstract}


\end{frontmatter}



\section{Introduction}
\label{sec:intro}

The chemical freeze-out marks an instant during the course of a heavy-ion collision at which the chemical composition of the emerging hadronic matter is fixed. The associated parameters, freeze-out temperature $T_{ch}$ and baryo-chemical potential $\mu_{B,ch}$, provide important information about the thermal conditions present at this stage. Commonly, these parameters are determined in dependence of the beam energy $\sqrt{s}$ by comparing measured particle ratios with statistical (thermal) hadronization model (SHM) calculations, cf.~e.g.~\cite{BraunMunzinger:2003zd,Cleymans:2005xv,Becattini:2005xt,Andronic:2011yq}. As an alternative, fluctuations in the conserved charges of QCD, e.g. net-baryon number $N_B$ and net-electric charge $N_Q$, were proposed to probe the freeze-out conditions in heavy-ion collisions~\cite{Karsch:2010ck,Karsch:2012wm}. 
The latter are expected to be negligibly affected by final state effects because the time scales for changing the corresponding local densities via diffusion processes are large. 

Experimentally, fluctuations in the conserved charges can only be studied by looking into a restricted kinematic acceptance region. Measurements of event-by-event fluctuations in the net-electric charge and net-proton number were recently reported in~\cite{Adamczyk:2013dal,Adamczyk:2014fia}. In contrast to $N_B$, however, the net-proton number is not a conserved quantity such that their fluctuations are affected by final state effects~\cite{Kitazawa:2012at}. From early measurements, various lattice QCD studies~\cite{Bazavov:2012vg,Borsanyi:2013hza} attempted to determine the freeze-out parameters in the $\mu_B$-$T$ plane. In a recent work~\cite{Borsanyi:2014ewa}, they were successfully extracted from the data~\cite{Adamczyk:2013dal,Adamczyk:2014fia} for the first time. Apart from other restrictions in comparison to the experimental situation lattice QCD approaches are, however, presently limited to a small region in $\mu_B$. 

Thermal models, instead, can be applied for arbitrary values of $\mu_B$ and allow to include various details of the experimental analysis. For a reasonable extraction of freeze-out conditions with thermal models it is necessary, however, that the fluctuations originate from a primordially equilibrated source and that the influence of critical fluctuations in the considered observables is negligible. Such an attempt was recently reported in~\cite{Alba:2014eba}, where freeze-out parameters have been obtained from a combined analysis of net-electric charge and net-proton number fluctuations. 

\section{Freeze-out conditions from net-electric charge and net-proton fluctuations}
\label{sec:conditions}

\begin{figure}
\begin{center}
\includegraphics*[scale=0.32]{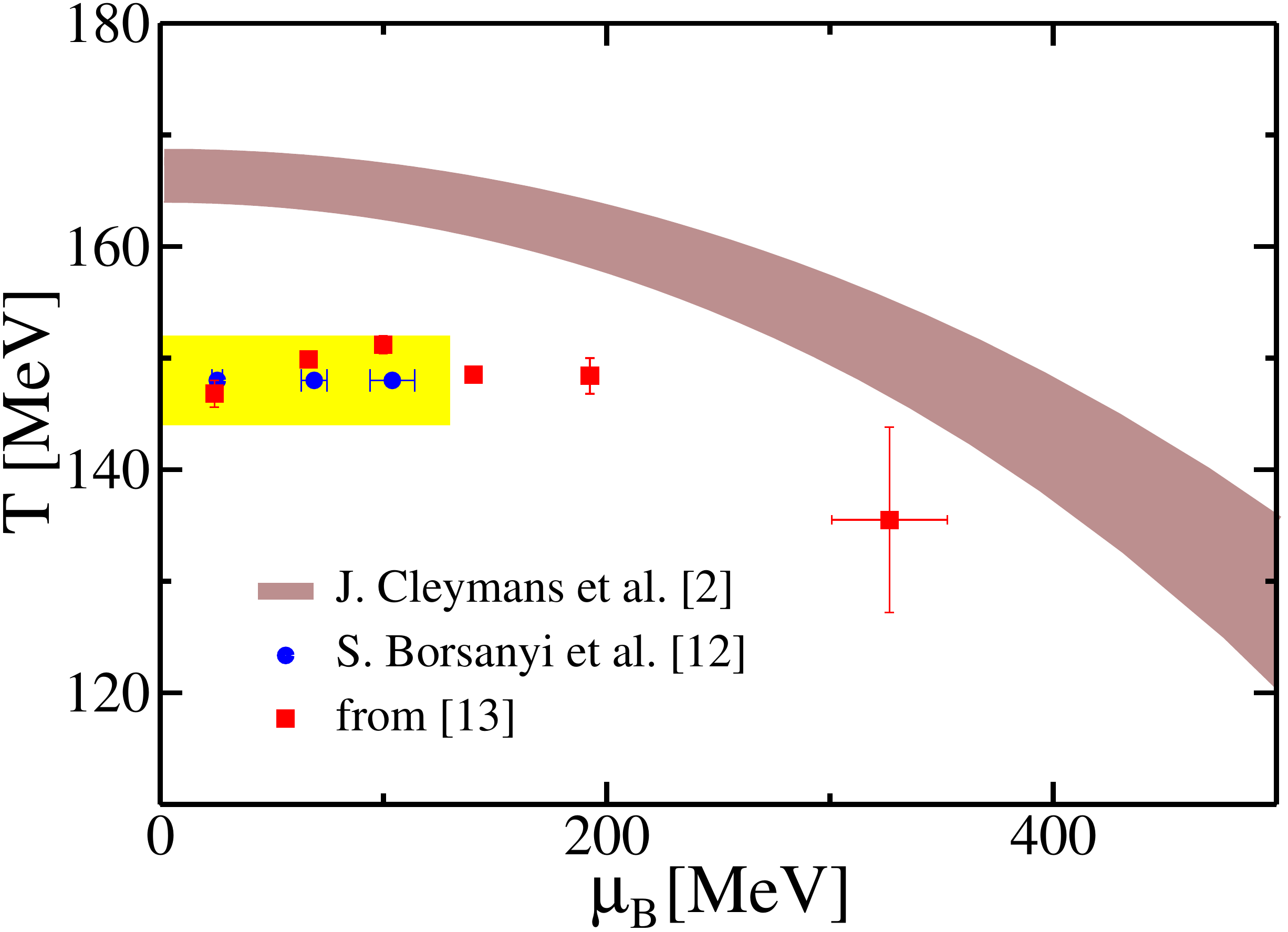}
\hspace{3mm}
\includegraphics*[scale=0.32]{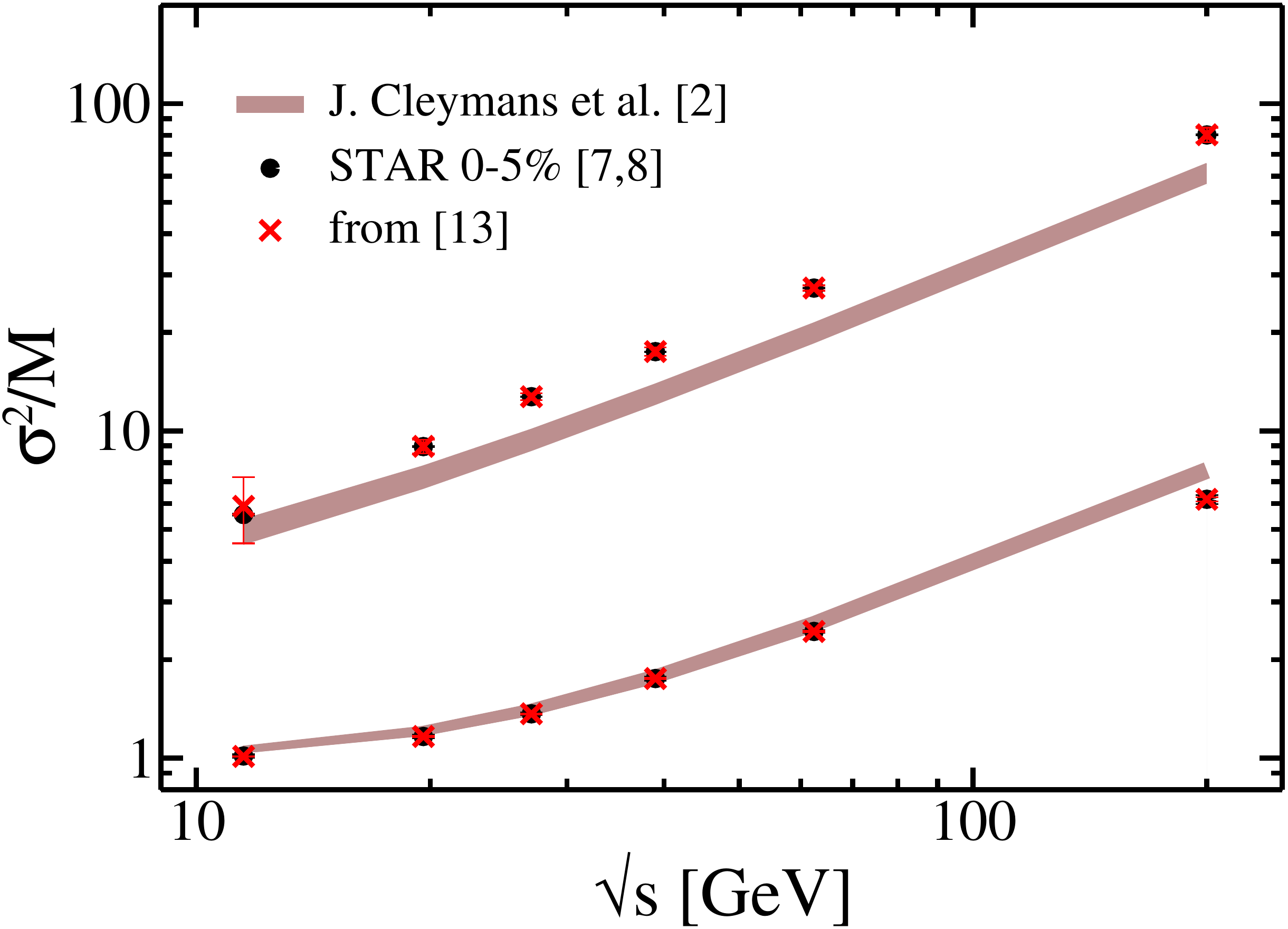}
\caption{(Color online) Left: comparison of freeze-out parameters $(\mu_{B,ch},T_{ch})$ in the $\mu_B$-$T$ plane. The curved band depicts a summary of freeze-out conditions obtained from various SHM fits, cf.~\cite{Cleymans:2005xv}. The horizontal band together with the circles are the lattice QCD results from~\cite{Borsanyi:2014ewa}: the band indicates that for $T_{ch}$ an upper limit $T_{ch}\leq(148\pm 4)$~MeV was determined, while the circles depict the corresponding mean value with error bars in $\mu_{B,ch}$. The squares show our freeze-out conditions found in~\cite{Alba:2014eba}. Right: comparison of the experimental data with model results for the cumulant ratio $\sigma^2/M$ of net-electric charge (top) and net protons (bottom) as a function of $\sqrt{s}$: 
the circles show the efficiency corrected STAR data~\cite{Adamczyk:2013dal,Adamczyk:2014fia} for most central collisions and the crosses our results obtained with the freeze-out parameters in~\cite{Alba:2014eba}. The bands depict the corresponding results found with the conditions~\cite{Cleymans:2005xv}.}
\label{fig:fig1}
\end{center}
\end{figure}
The statistical moments commonly determined from a measured multiplicity distribution are mean $M$, variance $\sigma^2$, skewness $S$ and kurtosis $\kappa$. They are related to the (net-)number $N$ of interest by $M=\langle N\rangle$, $\sigma^2=\langle(\Delta N)^2\rangle$, $S=\langle(\Delta N)^3\rangle/\sigma^3$ and $\kappa=\langle(\Delta N)^4\rangle/\sigma^4-3$, where $\Delta N=N-\langle N\rangle$ is the fluctuation of $N$ around its mean value. The cumulants $C_n$ of the distribution are defined as $C_1=M$, $C_2=\sigma^2$, $C_3=S\sigma^3$ and $C_4=\kappa\sigma^4$ and are, for an equilibrated system, related to generalized susceptibilities given by appropriate derivatives of the thermodynamic potential. 

In~\cite{Alba:2014eba}, we determined $T_{ch}$ and $\mu_{B,ch}$ in dependence of $\sqrt{s}$ by analyzing the efficiency-corrected\footnote{Efficiency corrected data for $\sigma^2/M$ for net protons can be found on the public STAR webpage.} STAR data~\cite{Adamczyk:2013dal,Adamczyk:2014fia} for $\sigma^2/M$ of the net-electric charge and net-proton multiplicity distributions for most central collisions with a thermal model in the grand canonical ensemble. The associated electric charge and strangeness chemical potentials were obtained by imposing conditions which reflect the physical situation in the initial state. In our model, we included hadrons and resonances with masses up to $2$~GeV$/$c listed by the Particle Data Group~\cite{Beringer:1900zz}. 
The experimentally employed kinematic cuts were applied in our approach on the level of primordial hadrons and resonances. In the analysis of net-electric charge fluctuations we included contributions from $\pi^+$, $\pi^-$, $K^+$, $K^-$, $p$ and $\bar{p}$. This system represents a suitable proxy for $N_Q$. The influence of resonance decays was also taken into account, where feed-down contributions from weak decays were excluded in agreement with the measurements. Furthermore, in the net-proton analysis we took the influence of isospin randomizing final state interactions into account, as explained in more detail in~\cite{Nahrgang:2014fza}. This turned out to be essential for a successful determination of common freeze-out conditions from combined net-electric charge and net-proton number fluctuations. 

The freeze-out conditions obtained in the analysis~\cite{Alba:2014eba} are shown in Fig.~\ref{fig:fig1} (left panel) together with other results found from SHM fits~\cite{Cleymans:2005xv} and in the recent lattice QCD study~\cite{Borsanyi:2014ewa}. The quality of the description of the STAR data~\cite{Adamczyk:2013dal,Adamczyk:2014fia} for $\sigma^2/M$ of the net-electric charge and net-proton distributions is compared for the two different parameter sets from~\cite{Cleymans:2005xv} and~\cite{Alba:2014eba} in the right panel of Fig.~\ref{fig:fig1}. Clearly, the measured lowest-order cumulant ratios are insufficiently reproduced for the values from the SHM fits~\cite{Cleymans:2005xv}. The smallness of the error bars in our extracted freeze-out parameters is a direct consequence of the small error bars in the STAR data, which are within the size of the symbols shown in Fig.~\ref{fig:fig1} (right panel). 

\section{Discussion and comments}
\label{sec:discussion}

For small $\mu_B$, the freeze-out points $(\mu_{B,ch},T_{ch})$ from~\cite{Alba:2014eba} are located at the lower edge of the confinement transition band determined in lattice QCD~\cite{Karsch:2013fga}. Furthermore, they agree remarkably well with the results of the lattice study in~\cite{Borsanyi:2014ewa}. In the latter, the STAR data~\cite{Adamczyk:2013dal,Adamczyk:2014fia} for different cumulant ratios of the net-electric charge and net-proton distributions were compared with corresponding susceptibility ratios of the net-electric charge and net-baryon number. 
Although the cumulant ratios in the net-baryon and net-proton numbers are, in principle, different observables they are related to each other when a binomial distribution for (anti-)protons among the (anti-)baryons can be assumed~\cite{Kitazawa:2012at}. It is interesting to note that in this case thermal model calculations show an approximate equivalence between the net-proton and net-baryon cumulant ratios~\cite{Nahrgang:2014fza}. 

The largest deviations in the freeze-out conditions from~\cite{Cleymans:2005xv} and~\cite{Alba:2014eba} are found for high beam energies, i.e.~small $\mu_B$. In the following, we want to comment on this observation and discuss mostly for $\sqrt{s}=200$~GeV the possible influence of additional effects neglected in our original analysis~\cite{Alba:2014eba}: 
\begin{enumerate}[(i)]
 \item A comprehensive comparison of different freeze-out conditions should include the investigation of particle ratios in addition to the cumulant ratios. Measured particle ratios for $\sqrt{s}=200$~GeV, which according to the STAR Collaboration are properly feed-down corrected for weak decays, can be found in~\cite{Andronic:2012dm}. Thermal model results obtained for the freeze-out conditions $(24.3,146)$~MeV were shown in~\cite{Alba:2014eba} and found to yield an overall equivalently good description of the particle ratios compared with the conditions $(24.3,166)$~MeV as contained in~\cite{Cleymans:2005xv}. 
 In fact, the parameters in~\cite{Alba:2014eba} describe (anti-)proton to pion ratios better but are worse in the description of ratios containing (multi-)strange hyperons. This suggests, together with the conclusions drawn from the right panel of Fig.~\ref{fig:fig1}, that our fluctuation observable analysis in~\cite{Alba:2014eba} is dominated by particles carrying light quark degrees of freedom. 
 \item For a complete randomization of isospin due to final state interactions both the pion density must be large and the duration of the hadronic stage long enough. In~\cite{Kitazawa:2012at} it was argued that these conditions are satisfied for beam energies $\sqrt{s}\geq 10$~GeV. Assuming that for the lower $\sqrt{s}$ the isospin is not fully randomized would result in a small decrease in $\sigma^2/M$ but a significant decrease in the higher-order cumulant ratios, cf.~\cite{Nahrgang:2014fza}. 
 \item The radial flow of the expanding matter is, in general, expected to influence the net-electric charge cumulant ratios due to the different masses of involved particle species. Using estimates for the expansion velocity based on STAR blastwave fits we found, however, that the impact on $\sigma^2/M$ for all $\sqrt{s}$ is small and within the error bars shown in Fig.~\ref{fig:fig1}. Correspondingly, the effect of radial flow on our extracted freeze-out parameters is negligible. This can be attributed to the rather large accepted transverse momentum window of $0.2$~GeV$/$c$\,\leq p_T\leq 2$~GeV$/$c in~\cite{Adamczyk:2014fia}. 
 \item The charges $N_B$ and $N_Q$ are conserved not only on average, as for a grand canonical ensemble, but event-by-event. The influence of exact (local) net-baryon number conservation on the net-proton cumulants was investigated in~\cite{Schuster:2009jv,Bzdak:2012an} and found to increase with decreasing beam energy and increasing order of the considered cumulant. Although it is hoped that a situation close to a grand canonical ensemble is realized experimentally through the application of kinematic cuts, the impact of potential deviations thereof may be estimated in line with~\cite{Bzdak:2012an}. 
 For $\sqrt{s}=200$~GeV, we find a reduction in $\sigma^2/M$ for net protons of about $1\%$ which is within the experimental error bars. The influence of exact baryon number conservation on our extracted freeze-out conditions is, therefore, minor for the higher $\sqrt{s}$. The impact of exact net-electric charge conservation is expected to be even smaller due to the large number of charged particles present at high beam energies~\cite{Bzdak:2012an}. 
 \item In the cumulant ratio $\sigma^2/M$, the system volume $V$ cancels out only to leading order. In general, volume fluctuations triggered by variations in the event-by-event geometry can affect fluctuations in the conserved charges. The possible influence on the net-baryon number cumulants was studied in~\cite{Skokov:2012ds}. It was found that corrections in $\sigma^2/M$ arising from volume fluctuations increase the cumulant ratio proportionally to the variance $\langle(\Delta V)^2\rangle$ in $V$. Based on Glauber Monte Carlo simulations the latter is found, however, to be numerically small for most central collisions~\cite{Skokov:2012ds}, such that we expect the impact of volume fluctuations on our results to be small. 
 \item In our approach, we considered point-like hadrons and resonances. The impact of excluded volume corrections on fluctuation observables was studied e.g.~in~\cite{Gorenstein:2007ep,Fu:2013gga}. In the case of net-proton number fluctuations, an increasing effect with decreasing $\sqrt{s}$ and increasing cumulant order was observed, where a negligible impact on $\sigma^2/M$ for all beam energies was found~\cite{Fu:2013gga}. It will be worthwhile to study the influence of this effect on the net-electric charge fluctuations in our analysis of freeze-out conditions in more detail in a future work. 
 \item Critical fluctuations can significantly enhance the fluctuations of thermal origin. They may arise either from the influence of a nearby critical point in the QCD phase diagram or, already for $\mu_B=0$, from a remnant criticality of the chiral phase transition in massless QCD. In the latter case, significant deviations in the net-baryon number (and net-electric charge) cumulants from thermal behavior are expected only for order $n\geq 6$ at vanishing or small $\mu_B$~\cite{Friman:2011pf}. In contrast, critical fluctuations associated with the second order phase transition at the conjectured critical point would alter already $\sigma^2$~\cite{Karsch:2010ck}. 
 Our determination of freeze-out conditions in~\cite{Alba:2014eba} is based on an analysis of data for $\sigma^2/M$. If critical fluctuations were present in these data, significant deviations from Poissonian behavior should be expected. For this reason, we investigated the ratio $\sigma^2_{p-\bar{p}}/(M_p+M_{\bar{p}})$ as a function of $\sqrt{s}$ and found no evidence for such a signal. It remains to be understood if the deviations of the data from thermal expectations seen in the higher-order cumulant ratios at lower $\sqrt{s}$~\cite{Alba:2014eba} could be of critical origin. 
\end{enumerate}

\section{Conclusions}
\label{sec:conclusion}

We discussed freeze-out conditions determined with a thermal model from fluctuation observables measured at RHIC and compared with corresponding lattice QCD and SHM fit results. While our results agree well with a recent lattice QCD study, major deviations from the freeze-out parameters deduced via SHM fits to particle ratios are found. The possible impact of various critical and non-critical fluctuation sources not contained in our original analysis was investigated. They affect our results negligibly for high $\sqrt{s}$. 

\section*{Acknowledgements}
The work is funded by the Italian Ministry of Education, Research and Universities under the FIRB Research Grant FIRB RBFR0814TT, a fellowship within the Postdoc-Program of the German Academic Exchange Service (DAAD), and the US Department of Energy grants DE-FG02-03ER41260, DE-FG02-05ER41367 and DE-FG02-07ER41521. 








\end{document}